\documentclass[11pt, letterpaper]{article}

\usepackage{amsmath}
\usepackage{amsfonts}
\usepackage{amssymb}
\usepackage{graphicx}
\usepackage{simplewick}

\pdfoutput=1

\usepackage{color}

\usepackage{amsmath, amssymb, graphics, epsfig, graphicx}
\usepackage{epsf}
\usepackage{epstopdf}
\usepackage {amssymb}
\newcommand{\nc}{\newcommand}
\nc{\ba}{\begin{eqnarray}}
\nc{\ea}{\end{eqnarray}}
\newcommand\be{\begin{equation}}
\newcommand\ee{\end{equation}}

\newcommand{\calR}{{\cal{R}}}

\newcommand{\calP}{{\cal{P}}}

\newcommand{\bea}{\begin{eqnarray}}
\newcommand{\eea}{\end{eqnarray}}

\newcommand{\bfx}{{\bf{x}}}
\newcommand{\bfq}{{\bf{q}}}
\newcommand{\bfk}{{\bf{k}}}
\newcommand{\bfp}{{\bf{p}}}

\begin{document}

\vspace{5mm}
\vspace{0.5cm}
\begin{center}

\def\thefootnote{\fnsymbol{footnote}}

{ \bf  \large Loop Corrections in  Bispectrum in USR  Inflation with PBHs Formation }
\\[1cm]

{ Hassan Firouzjahi$\footnote{firouz@ipm.ir}$}\\[0.5cm]

{\small \textit{ School of Astronomy, Institute for Research in Fundamental Sciences (IPM) \\ P.~O.~Box 19395-5531, Tehran, Iran }}\\

\end{center}

\vspace{.8cm}

\hrule \vspace{0.3cm}


\begin{abstract}

We calculate the one-loop corrections in bispectrum of CMB scale perturbations induced from the small scale modes undergoing an intermediate phase of USR inflation in scenarios employed for PBHs formation. Using the formalism of effective field theory of inflation  we calculate the cubic and quartic Hamiltonians and perform the in-in analysis  for a subset of Feynman diagrams comprising both the cubic and the quartic exchange vertices.  We show the one-loop corrections in bispectrum has 
the local shape with $f_{NL}$ having the same structure as the   one-loop correction in power spectrum in their  dependence on the duration of the USR phase and the sharpness of the transition  to the final attractor phase. It is shown that  in the models with a sharp transition 
the induced loop corrections in bispectrum  can quickly violate the observational bounds on $f_{NL}$. 

\end{abstract}
\vspace{0.5cm} \hrule
\def\thefootnote{\arabic{footnote}}
\setcounter{footnote}{0}
\newpage
\section{Introduction}
\label{intro}

Models of inflation incorporating an intermediate phase of ultra slow-roll (USR) inflation have been actively employed  in generating  primordial black holes (PBHs) as a candidate for dark matter \cite{Ivanov:1994pa, Garcia-Bellido:2017mdw, Germani:2017bcs, Biagetti:2018pjj}, for a review see \cite{Khlopov:2008qy, Ozsoy:2023ryl, Byrnes:2021jka}. More specifically, during the USR phase of inflation with a flat potential, the curvature perturbation power spectrum grows on superhorizon scales 
so it can be enhanced significantly compared to the long CMB scales to source the PBHs formation. On the other hand, the rapid rise of the curvature perturbation power spectrum during the USR phase may cause troubles. Indeed, it was argued in  \cite{Kristiano:2022maq, Kristiano:2023scm} that the  one-loop corrections\footnote{For  earlier works on loop corrections in power spectrum during  inflation see for example \cite{ Seery:2007wf, Seery:2007we, Senatore:2009cf, Pimentel:2012tw, Inomata:2022yte}.}  from small scales USR modes  can significantly affect the long CMB scale perturbations. It was argued in  \cite{Kristiano:2022maq, Kristiano:2023scm} that the model is not trusted to generate the  desired PBHs abundance as it is not perturbatively under control.  
Thereafter, the question of one-loop corrections in power spectrum in these models  has attracted considerable interests \cite{ Riotto:2023hoz, Riotto:2023gpm, Choudhury:2023vuj,  Choudhury:2023jlt,  Choudhury:2023rks, Choudhury:2023hvf, 
Choudhury:2024one, Firouzjahi:2023aum, Motohashi:2023syh, Firouzjahi:2023ahg, Tasinato:2023ukp, Franciolini:2023lgy, Firouzjahi:2023btw, Maity:2023qzw, Cheng:2023ikq, Fumagalli:2023loc, Nassiri-Rad:2023asg, Meng:2022ixx, Cheng:2021lif, Fumagalli:2023hpa, Tada:2023rgp, Firouzjahi:2023bkt, Iacconi:2023slv}.
For example,  the conclusion in \cite{Kristiano:2022maq, Kristiano:2023scm} was 
criticized in  \cite{ Riotto:2023gpm, Riotto:2023hoz} where it was argued that  the dangerous one-loop corrections can be harmless in a smooth transition. This question was further studied in  \cite{Firouzjahi:2023ahg} where,  employing $\delta N$ formalism,  it was shown that for a mild transition the one-loop corrections are suppressed by the slow-roll parameters so the setup is reliable  for PBHs formations. This question was also studied numerically in \cite{Davies:2023hhn}
and using the separate universe  formalism in \cite{Iacconi:2023ggt}. 

To calculate the full one-loop corrections in curvature perturbation power spectrum, one needs to incorporate the effects of both cubic and quartic interactions. While the cubic interactions can be borrowed from \cite{Maldacena:2002vr} but the situation for the quartic interactions is somewhat difficult as calculating the quartic action in this setup is non-trivial, for earlier studies on quartic action see \cite{Jarnhus:2007ia, Arroja:2008ga}.  
In \cite{Firouzjahi:2023aum}, employing the formalism of effective field theory (EFT) of inflation, we have studied this question in which the effects of both cubic and quartic Hamiltonians were included. In addition, the effects of the sharpness of the transition from the intermediate USR phase to the final attractor phase were studied as well. It was shown in \cite{Firouzjahi:2023aum} that  loop corrections 
from the quartic Hamiltonian are comparable to the loop corrections from the cubic interaction. The analysis of \cite{Firouzjahi:2023aum} supports the conclusion of  \cite{Kristiano:2022maq} that the loop corrections can be significant 
for the setup  where the transition from the USR phase to the final attractor phase is sharp while the loop corrections can be washed out during a mild transition. 

It may look counterintuitive, based on the notion of decoupling of scales,  
as how small scales can affect the long modes. This question was originally reviewed in  \cite{ Riotto:2023gpm} and further studied in  \cite{Firouzjahi:2023bkt}. The basic idea is that the non-linear coupling between the long and short modes  provides the source term for the evolution of the long mode. On the other hand,  the long mode modulates the spectrum of the short modes. This modulation becomes significant if the power spectrum of the short modes experiences a significant scale-dependent enhancement as in USR setup.  The combination of the non-linear coupling between the  long and short modes and the modulation of the short modes by the long mode back-reacts on the long mode itself and induces the  one-loop correction \cite{Riotto:2023hoz, Franciolini:2023lgy, Firouzjahi:2023bkt}.

Motivated by the question of one-loop corrections in power spectrum, it is a natural question to look for the one-loop corrections in bispectrum on CMB scales. Indeed, in the single field slow-roll (SR) setups, the non-Gaussianity parameter $f_{NL}$ is very small  \cite{Maldacena:2002vr} so the long modes which leave the horizon during the early SR phase have negligible level of non-Gaussianity. On the other hand, if the loop corrections from small USR scales on large scale 
power spectrum are significant, then it is natural that the one-loop corrections in bispectrum to be significant as well. This is a non-trivial question as the corresponding  analysis requires the cubic, quartic and the quintic interaction Hamiltonians involving more complicated in-in analysis compared to the case of loop corrections in power spectrum. This is the main goal of this paper in which, 
using the EFT formalism of inflation as in \cite{Firouzjahi:2023aum}, we calculate the one-loop corrections in bispectrum on CMB scales.


\section{The Setup}
\label{setup}

In this section we present our setup. As  in \cite{Kristiano:2022maq, Kristiano:2023scm}, it is a single field model of inflation containing three stages $SR \rightarrow USR \rightarrow SR$. The first and and the third stages are SR phases while the intermediate stage is a USR phase. The large scale CMB perturbations leave the horizon during the first stage with an amplitude of curvature perturbation fixed by the COBE normalization. 
The second phase is tuned to generate the PBHs at the desired scale consistent with various cosmological observations. Typically, the intermediate USR phase starts around 30 e-folds after the long CMB modes leave the horizon and it lasts for about 2-3 e-folds. The USR phase is followed by the third SR phase in which the system reaches its final attractor phase. 

During the SR phases the curvature perturbation $\calR$ is frozen on superhorizon scales and the amplitude of non-Gaussianity parameter $f_{NL}$ is slow-roll suppressed. In addition, there is a consistency condition  between the two-point  and the three-point correlation functions as shown by 
Maldacena \cite{Maldacena:2002vr, Creminelli:2004yq}. On the other hand, the USR setup is a single field model of inflation in which the potential is flat \cite{Kinney:2005vj, Morse:2018kda, Lin:2019fcz, Dimopoulos:2017ged} so the inflaton velocity falls off exponentially and the curvature perturbations grow on superhorizon scales \cite{Namjoo:2012aa}. Since the curvature perturbation is not frozen on superhorizon scales in USR setup, it provides a counterexample to violate the  Maldacena consistency condition \cite{Namjoo:2012aa, Martin:2012pe, Chen:2013aj, Chen:2013eea, Akhshik:2015nfa, Akhshik:2015rwa, Mooij:2015yka, Bravo:2017wyw, Finelli:2017fml, Passaglia:2018ixg, Pi:2022ysn, Ozsoy:2021pws,  Firouzjahi:2023xke, Namjoo:2023rhq, Namjoo:2024ufv}. 
The amplitude of the local-type non-Gaussianity in conventional USR model in which the USR phase is sharply followed by an attractor SR phase 
is  $f_{NL}=\frac{5}{2}$  \cite{Namjoo:2012aa}. However, it was shown in \cite{Cai:2018dkf} that the final amplitude of $f_{NL}$  depends on the sharpness of the transition from the USR phase to the final SR phase. In particular,  for a mild transition the curvature perturbations evolve after the USR phase until it reaches to its final attractor value. Because of this evolution,  much of the amplitude of $f_{NL}$ is washed out towards the end of inflation. The important lesson from this study is that   the sharpness of the transition from the USR phase to the final SR phase plays important roles when measuring the cosmological observables at the time of end of inflation.  

 Starting with the FLRW metric
\ba
ds^2 = -dt^2 + a(t)^2 d{\bf x}^2 \, ,
\ea
the background fields equations for the inflaton field $\phi$ and the scale factor $a(t)$ are  given by
\ba
\ddot \phi(t) + 3 H \dot \phi(t)=0\, , \quad \quad 3 M_P^2 H^2 \simeq V_0, 
\ea
in which $M_P$ is the reduced Planck mass, $H$ is the Hubble expansion  rate during inflation and $V_0$ is the value of the potential during the USR phase which is constant. Consequently,  $H$ is nearly constant while 
$\dot \phi \propto a^{-3}$ during the USR phase.  

The slow-roll  parameters related to  $H$  are  defined as usual,
\ba
\label{ep-eta}
\epsilon \equiv -\frac{\dot H}{H^2} =\frac{\dot \phi^2}{2 M_P^2 H^2}\, , \quad \quad 
\eta \equiv \frac{\dot \epsilon}{H \epsilon} \, . 
\ea
During the  SR phases  both $\epsilon$ and $\eta$ are nearly constant and small but during the USR phase $\epsilon$ falls off like $a^{-6}$ 
while $\eta\simeq -6$. Going to conformal time $d \tau= dt/a(t)$ with $a H \tau \simeq -1$, $\epsilon(\tau)$ scales with time  as 
 \ba
 \epsilon(\tau) = \epsilon_i \big( \frac{\tau}{\tau_s} \big)^6 \, ,
 \ea
in which $\epsilon_i$ is the value of $\epsilon$  during the first SR phase which is nearly constant.  
We assume the USR phase is extended between the interval $\tau_s < \tau <\tau_e$ so  $\epsilon$ at the end of USR phase is $\epsilon_e = \epsilon_i \big( \frac{\tau_e}{\tau_s} \big)^6 $. Defining the number of e-fold as $d N= H dt$, the duration of the USR phase is given by $\Delta N \equiv N(\tau_e) - N(\tau_s)$ 
yielding to    $\epsilon_e = \epsilon_i e^{-6 \Delta N}  $. 

An important question in this study is how the USR phase is glued to the final attractor phase.  As in \cite{Cai:2018dkf}, we assume the potential in the final SR phase has the following form 
\ba
V(\phi) = V(\phi_e) + \sqrt{2 \epsilon_V} V(\phi_e)  (\phi -\phi_e) + \frac{\eta_V}{2} V(\phi_e) (\phi -\phi_e)^2 + ... \, ,
\ea
in which $2\epsilon_V \equiv  M_P^2\big(V'(\phi_e)/V(\phi_e) \big)^2$ and $\eta_V\equiv  M_P^2 V''(\phi_e)/V(\phi_e)$ are the  slow-roll parameters 
defined in terms of the  derivatives of the potential. We assume that 
the potential is continuous at $\phi=\phi_e$ but its derivative may not be continuous with $\epsilon_V\neq0$ so there is a kink in the potential.  To simplify the setup, we assume $\eta_V=0$ and the transition to the final stage is sharp. However, this is not a restrictive assumption and our analysis can be extended to the case where $\eta_V\neq0$ as well.

Let us set $N=0$ to be the time of the transition from the USR phase to the final SR phase. Solving the background field equation in the final SR phase, and imposing the continuity of $\phi$ and $\frac{d \phi}{d N}$  at the time of transition  we obtain \cite{Cai:2018dkf} (see also \cite{Cai:2022erk, Kawaguchi:2023mgk})
\ba
\label{phi-back}
M_P^{-1}\phi(N)= \frac{C_1}{3} e^{-3 N} + \frac{h}{6}  \sqrt{2 \epsilon_V} N + C_2 \, ,
\ea
with the constants of integration $C_1$ and $C_2$  given by
 \ba
 C_1=  \sqrt{2 \epsilon_e} (1 + \frac{h}{6} ) \, , \quad \quad 
 C_2 = M_P^{-1} \phi_e - \frac{ \sqrt{2 \epsilon_e}}{3} (1 + \frac{h}{6} ) \, ,
 \ea
in which, following \cite{Cai:2018dkf}, we have  defined the sharpness parameter $h$ via
\ba
\label{h-def}
h\equiv \frac{6 \sqrt{2 \epsilon_V} }{\dot \phi(t_e)} M_P = -6 \sqrt{\frac{\epsilon_V}{\epsilon_e}} \, .
\ea
Since we work with the convention that  $\phi$ is decreasing monotonically during inflation, then $\dot \phi<0$ and $h<0$.  As shown in \cite{Cai:2018dkf}, $h$ is the key parameter of the setup,  controlling the sharpness of the transition from the USR phase to the final SR phase. 

With the background dynamics given as in Eq. (\ref{phi-back}), the SR parameters defined in Eq. (\ref{ep-eta}) for the final SR phase $(N>0)$ are given by
\ba
\label{ep-N}
\epsilon(\tau)= \epsilon_e  \Big(\frac{h}{6} - (1+ \frac{h}{6} ) \big(\frac{\tau}{\tau_e} \big)^3 \Big)^{2} \, ,
\ea
and
\ba
\label{eta-N}
\eta(\tau) = -\frac{6 (6+h)}{(6+h) - h   \big(\frac{\tau_e}{\tau} \big)^3} \, .
\ea
Towards the final stage of inflation when $\tau \rightarrow \tau_0 \rightarrow 0$,  $\epsilon \rightarrow \epsilon_e (\frac{h}{6})^2$ and $\eta $ vanishes like $\tau^3$. While $\epsilon$ is  smooth  across the transition point but it is important to note that $\eta$ has a jump  at $\tau=\tau_e$. More specifically, just prior to the transition (i.e. near the end of USR phase) $\eta=-6$ while right after the transition  $\eta= -6-h$. As a result, near the transition point one can approximate $\eta$ via \cite{Cai:2018dkf}
\ba
\eta = -6 - h \theta(\tau -\tau_e) \quad \quad  \tau_e^- < \tau < \tau_e^+ \, .
\ea
With this approximation, we obtain   
\ba
\label{eta-jump}
\frac{d \eta}{d \tau} = - h \delta (\tau -\tau_e)  \, ,  \quad \quad  \tau_e^- < \tau < \tau_e^+ \, .
\ea

For an infinitely sharp transition $h \rightarrow -\infty$ so  after the transition $\epsilon$ evolves rapidly to a larger value such that $\epsilon$ at the end of inflation 
  is given by $\epsilon(\tau_0) \simeq \epsilon_V = \epsilon_e (\frac{h}{6})^2$. 
For an ``instant" sharp transition which was assumed  in \cite{Kristiano:2022maq, Kristiano:2023scm} we have $h=-6$.  In this limit  $\epsilon$ in the final SR phase is frozen to $\epsilon_e$ fixed at the end of USR phase.

To implement the in-in formalism we need the mode function for comoving curvature perturbation  $\calR$ during the USR and the follow up  SR phase. Going to Fourier space,  the mode function  is written as
\begin{equation}
\calR({\bf x}, t) = \int \frac{d^3 k}{(2\pi)^3} e^{i {\bf k}\cdot {\bf x}} \hat\calR_{\bf k}(t) \, ,
\end{equation}
 in which the operator $\hat\calR_{\bf k}(t)$ is expressed in terms of the creation and annihilation operators as $\hat\calR_{\bf k}(t)= \calR_k(t) a_{\bf k} + \calR^*_k(t) a_{-\bf k}^\dagger$. Note that $\hat\calR_{\bf k}$ is a quantum operator while 
 $\calR_k$ is the usual mode function.  The creation and annihilation operators satisfy the usual commutation relations $[ a_{\bf k}, a^\dagger_{-\bf k'} ] = ( 2 \pi)^3 \delta (  {\bf k} + {\bf k'}) $.

Starting with the  Bunch-Davies initial condition, the  mode function during the first stage for $\tau < \tau_i$ is given by
 \begin{equation}
\calR^{(1)}_{k} =  \frac{H}{ M_P\sqrt{4 \epsilon_i k^3}}
( 1+ i k \tau) e^{- i k \tau} \, , \quad \quad (\tau < \tau_s) \, .
\end{equation}
During the USR phase, the mode function is given  by
\begin{equation}
\calR^{(2)}_{k} =  \frac{H}{ M_P\sqrt{4 \epsilon_i k^3}}  \bigg( \frac{\tau_s}{\tau} \bigg)^3
\Big[ \alpha^{(2)}_k ( 1+ i k \tau) e^{- i k \tau}  + \beta^{(2)}_k ( 1- i k \tau) e^{ i k \tau}  \Big]  \, ,
\end{equation}
with the coefficients $\alpha^{(2)}_k$ and $\beta^{(2)}_k$ fixed by imposing the continuity of the mode function and its time derivative as follows: 
\begin{equation}
\label{alpha-beta2}
\alpha^{(2)}_k = 1 + \frac{3 i }{ 2 k^3 \tau_s^3} ( 1 + k^2 \tau_s^2) \, , \quad \quad
\beta^{(2)}_k= -\frac{3i }{ 2 k^3 \tau_s^3 } {( 1+ i k \tau_s)^2} e^{- 2 i k \tau_s} \, .
\end{equation}
Finally, imposing the matching conditions at $\tau_e$, the mode function in the final SR phase  is obtained to be \cite{Firouzjahi:2023aum}
\begin{equation}
\label{R3-mode}
\calR^{(3)}_{k} =  \frac{H}{ M_P\sqrt{4 \epsilon(\tau) k^3}}
\Big[ \alpha^{(3)}_k ( 1+ i k \tau) e^{- i k \tau}  + \beta^{(3)}_k ( 1- i k \tau) e^{ i k \tau}  \Big] \, ,
\end{equation}
with $\epsilon(\tau)$ given by Eq.~(\ref{ep-N}) and
the coefficients $\alpha^{(3)}_k$ and $\beta^{(3)}_k$ are given by,
$$
\label{alpha-beta3}
\alpha^{(3)}_k = \frac{1}{8 k^6 \tau_s^3 \tau_e^3}  \Big[ 3h
 ( 1 -i k \tau_e)^2 (1+i k \tau_s)^2 e^{2i k (\tau_e- \tau_s)}
-i (2 k^3 \tau_s^3 + 3i k^2 \tau_s^2 + 3 i) (4 i k^3 \tau_e^3- h k^2 \tau_e^2 - h) \Big],
\nonumber
$$
and
$$
\beta^{(3)}_k=   \frac{-1}{8 k^6 \tau_s^3 \tau_e^3}  \Big[ 3 ( 1+ i k \tau_s)^2 ( h+ h k^2 \tau_e^2 + 4 i k^3 \tau_e^3 ) e^{-2 i k \tau_s} + i h ( 1+ i k \tau_e)^2  ( 3 i + 3 i k^2 \tau_s^2 + 2 k^3 \tau_s^3 ) e^{- 2 i k \tau_e}
 \Big]. \nonumber
$$

With the above mode functions at hands, we are ready to calculate the one-loop corrections in bispectrum. To clarify the notation, the momentum for the
 CMB modes are denoted by  ${\bf p}_1$,  ${\bf p}_2$ and $\bfp_3$ while  the momentum for the small scale modes which run in the loop are 
 denoted by ${\bf q}$. We have the vast hierarchy $p_i \ll q$. 
 Since we consider the loop corrections from the amplified 
 modes which leave the horizon during the USR phase, we cut the loop integrals in the intervals $ q_s \leq q < q_e $ in which $q_s = -\frac{1}{\tau_s}$ and $q_e= - \frac{1}{\tau_e}$. In addition, the duration of  the USR phase  $\Delta N\equiv  N(\tau_e) - N(\tau_s)$ is related to $q_s$ and $q_e$ via
\begin{equation}
\label{delta-N-def}
e^{- \Delta N}=  \frac{\tau_e}{\tau_s} = \frac{q_s}{q_e}  \, .
\end{equation}
To generate PBHs formation with the desired properties, one typically requires 
$\Delta N \sim 2-3$.

Before closing this review about our setup, there is an important comment in order.
In our setup, we consider an instant transition at $\tau=\tau_e$ to the final SR phase. On the other hand, the mode functions may evolve after the transition until it reaches its attractor value. This is controlled by the sharpness parameter $h$. For example, for an instant sharp transition with $h =-6$ as studied in \cite{Kristiano:2022maq, Kristiano:2023scm}, the final value of $\calR$ at the end of inflation is by a factor of $1/4$ smaller than its value at the end of USR. This is because the mode function keeps evolving until it reaches its attractor value. On the other hand, for an extreme sharp transition with $h \rightarrow -\infty$, the mode function freezes immediately after the USR phase. This is the limit which was studied in \cite{Namjoo:2012aa} yielding to $f_{NL}=\frac{5}{2}$. However, as shown in  \cite{Cai:2018dkf}, for mild transition with $|h| \ll 1$, most of non-Gaussianity is washed out during the subsequent evolution of USR phase. 

With the above discussions in mind, we should distinguish between the  instant transition and the sharp transition. In our analysis above, we have assumed an instant transition but with different values of the 
sharpness parameter  $h$. In particular, one can also relax the assumption of an instant transition and assume the transition may not happen instantly at $\tau= \tau_e$ \cite{Franciolini:2023lgy, Davies:2023hhn}. In this case, the outgoing mode function will be more complicated than what is obtained in 
Eq. (\ref{R3-mode}).  This will bring more complexities in theoretical analysis which is beyond the scope of our current work.

\begin{figure}[t]
	\centering
	\includegraphics[ width=0.68\linewidth]{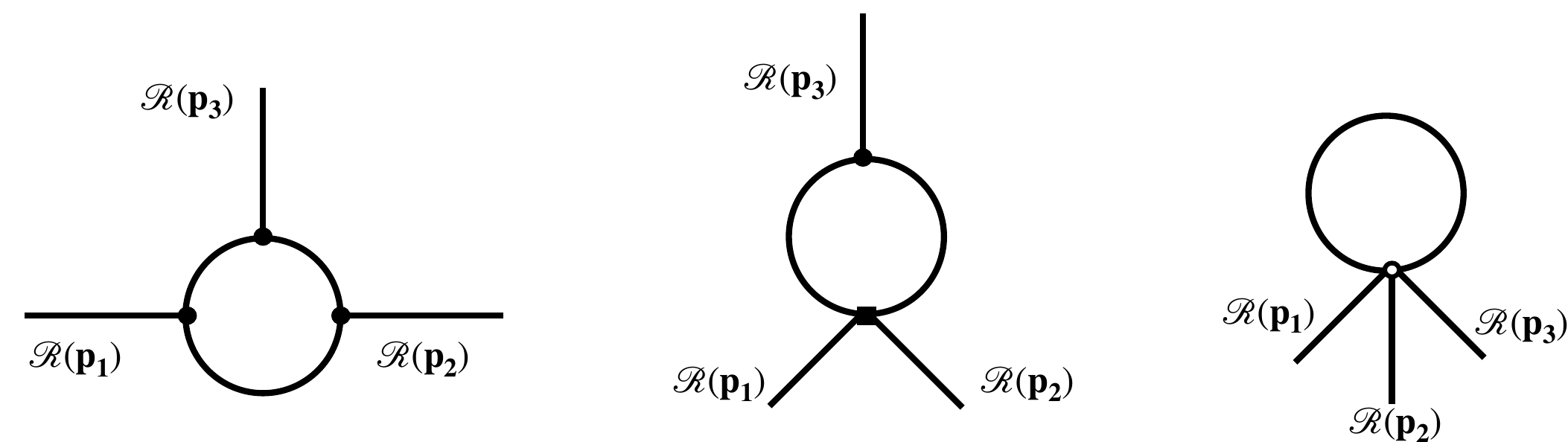}
	\vspace{.5 cm}
	\caption{ The one-particle irreducible Feynman diagrams for the one-loop corrections in bispectrum.  In the left panel all three vertices come from ${\bf H_3}$ while in the middle panel the lower vertex is from ${\bf H_4}$ and the upper one  is from ${\bf H_3}$. In the right panel there is a single vertex from ${\bf H_5}$. 
}
\label{Feynman-fig}
\end{figure}


\section{ Interaction Hamiltonians}

To calculate the one-loop corrections in bispectrum we need the interaction Hamiltonians. There are three different one-particle irreducible 
Feynman diagrams for the one-loop corrections in bispectrum as depicted in Fig. \ref{Feynman-fig}. The left panel represents the corrections entirely from cubic Hamiltonian ${\bf H}_3$.
The middle panel represents a mixed contributions in which the lower vertex comes from the quartic interaction Hamiltonian  ${\bf H}_4$ while the upper vertex comes from ${\bf H}_3$. Finally, the right panel represents the contribution from a single vertex containing  the quintic interaction Hamiltonian 
${\bf H}_5$. The contributions from the left diagram in Fig. \ref{Feynman-fig} is more complicated than the other two diagrams. This is because one has to consider a threefold time integrals in the in-in integrals. On the other hand, the in-in analysis associated with the middle diagram is somewhat easier as one deals with a twofold time integrals as we will demonstrate below. The in-in integral for the right panel involves only a single time integral so it is the easiest compared to other two diagrams. However, to calculate the contribution of the right diagram, we need to calculate the total ${\bf H}_5$. Besides the intrinsic quintic Hamiltonian constructed directly from  the action
in the form ${\bf L_5} \rightarrow -{\bf H}_5 $, the cubic and quartic interactions will induce  additional contributions in quintic interaction as well. 
This will bring more complexity into the analysis for the diagram in  the right panel of Fig. \ref{Feynman-fig}. In this work, as a first attempt to calculate the one-loop corrections in bispectrum, we consider the contribution of the middle diagram of Fig. \ref{Feynman-fig}.
The results of this analysis will provide useful information about the behaviour of the one-loop corrections. We postpone a complete study  
of the one-loop corrections in bispectrum involving the contributions of all three diagrams of Fig. \ref{Feynman-fig} to a future work.

The interaction Hamiltonians ${\bf H}_3$ and ${\bf H}_4$ for the USR setup was calculated in \cite{Firouzjahi:2023aum} employing the method of EFT of inflation \cite{Cheung:2007st, Cheung:2007sv} which we briefly review here. The EFT approach was originally employed in \cite{Akhshik:2015nfa} 
to calculate the tree-level bispectrum for a general $P(X, \phi)$-type of non-attractor setup. The EFT of inflation is a powerful tool in the decoupling limit when one neglects the gravitational back-reactions and  the matter perturbations comprise the dominant contributions.   In a near dS background with a time-dependent inflaton field $\phi(t)$, the four-dimensional diffeomorphism invariance  is spontaneously broken to a three-dimensional spatial diffeomorphism invariance. In  the unitary  gauge where the perturbations of inflaton are turned off
one  writes down all terms in the action which are consistent with the remaining three-dimensional diffeomorphism invariance. Correspondingly,  the background inflation dynamics is controlled via the  known  Hubble expansion rate $H(t)$ and its derivative $\dot H(t)$. After writing the action consistent with the three-dimensional diffeomorphism invariance,  one restores the full four-dimensional diffeomorphism invariance  by introducing a scalar field fluctuations, $\pi(x^{\mu})$, the Goldstone boson representing the breaking of  the time diffeomorphism invariance. As mentioned before, the advantage of the EFT approach is when one goes to the decoupling limit where the gravitational back-reactions are  slow-roll  suppressed 
and negligible. To calculate the interaction Hamiltonians, we have to expand the quantities $H(t + \pi)$ and $\dot H(t+\pi)$ to the corresponding orders. 
 
With the above discussions in mind the quadratic, cubic and quartic actions in the decoupling limit were calculated in \cite{Firouzjahi:2023aum}.  The quadratic action necessary to quantize the free theory is given by
\begin{equation}
\label{S2}
S_2= M_P^2 \int d\tau d^3 x   a^2 \epsilon H^2 \big(  \pi'^2 - (\partial_i \pi)^2 \big) \,,
\end{equation}
in which a prime denotes the derivative with respect to the conformal
time. 

The cubic action is obtained to be 
\begin{equation}
\label{action3}
S_{\pi^3} =  M_P^2 H^3 \int d\tau d^3 x\,  \eta \epsilon  a^2\,
\Big[  \pi \pi'^2  - \pi (\partial \pi)^2 \Big]  \, ,
\end{equation}
leading to the following  cubic interaction Hamiltonian, 
\begin{eqnarray}
\label{H3}
{\bf H}_3 =  
- M_P^2 H^3 \eta \epsilon a^2\, \int d^3 x  \Big[  \pi \pi'^2  +\frac{1}{2} \pi^2 \partial^2 \pi \Big] \, .
\end{eqnarray}
The above cubic Hamiltonian agrees  with that of~\cite{Akhshik:2015nfa} when $c_s=1$.  Note that the  gradient term can not be ignored a priori since its effects can be important \cite{Akhshik:2015nfa, Fumagalli:2023hpa}.

Similarly,  the quartic action is obtained to be 
\begin{equation}
\label{action4}
S_{\pi^4} =  \frac{M_P^2}{2} \int d\tau d^3 x\,   \epsilon a H^3
\big( \eta^2  a H + \eta' \big)\,    \Big[ \pi^2 \pi'^2 - \pi^2 (\partial \pi)^2 \Big] \, .
\end{equation}
Note the important contribution from the term $\eta' $ which induces a delta contribution in the interaction Hamiltonian when $\eta$ undergoes a jump from the USR phase to the third slow-roll phase as seen in Eq.~(\ref{eta-jump}).

As observed in  \cite{Firouzjahi:2023aum},  to calculate the quartic Hamiltonian we have to be careful as the time derivative interaction $\pi' \pi^2$ in ${\bf H}_3$ induces a new term in the quartic Hamiltonian~\cite{Chen:2006dfn, Chen:2009bc} so one can not simply conclude ${\bf H}_4 = -{\bf L}_4$.  More specifically, the quartic Hamiltonian receives additional  contribution  $+M_P^2 H^4  \eta^2 \epsilon a^2\,    \pi^2 \pi'^2$ from the cubic action, so the total quartic Hamiltonian is obtained to be
\begin{equation}
\label{H4}
{\bf H}_4 =  \frac{M_P^2}{2} \epsilon a H^3  \int d^3 x \Big[
\big(  \eta^2  a H - \eta'   \big)  \pi^2 \pi'^2
+ \big(  \eta^2  a H + \eta'  \big)  \pi^2 (\partial \pi)^2
\Big] \, .
\end{equation}
The interaction Hamiltonians (\ref{H3}) and (\ref{H4}) will be used in our analysis in the next section  to calculate the one-loop corrections in bispectrum for the middle diagram in Fig. \ref{Feynman-fig}. 


As discussed in \cite{Firouzjahi:2023aum} note that  we are interested in curvature perturbation on comoving surface $\calR$ while the above interaction Hamiltonians are written in terms of the variable $\pi$. There are additional contributions from the non-linear relations between $\pi$ and $\calR$. For example,  to cubic order in $\pi$, the non-linear relation between $\calR$ and $\pi$ is given by~\cite{Jarnhus:2007ia, Arroja:2008ga}
 \begin{eqnarray}
 \label{pi-R1}
 \calR = - H \pi + \Big( H \pi \dot \pi + \frac{\dot H}{2} \pi^2 \Big)
 + \Big( -H \pi \dot \pi^2 -\frac{H}{2} \ddot \pi \pi^2 - \dot H \dot \pi \pi^2 -\frac{\ddot H}{6} \pi^3 \Big) \, .
 \end{eqnarray}
 However, we calculate the three-point correlation function at the time of end of inflation $\tau=\tau_0 \rightarrow 0$ when the system is in the slow-roll regime and the perturbations are frozen on superhorizon scales, $\dot \pi =\ddot \pi=0$. Fortunately, in this limit all the non-linear corrections in $\calR$ in Eq.~(\ref{pi-R1}) are suppressed  so one can simply assume the  linear relation between $\calR$ 
 and $\pi$ 
 \begin{equation}
 \label{pi-R}
 \calR = - H \pi \, ,\quad  \quad \quad (\tau \rightarrow \tau_0).
 \end{equation}
Because of this linear relation between $\pi$ and $\calR$,  one can
use $\pi$ and $\calR$ interchangeably in the following in-in integrals with the mode function of $\calR$ from the free theory (in the interaction picture).

Before closing this section we comment about the roles of the total time derivative terms which caused controversies in recent literature, see for example \cite{Fumagalli:2023hpa, Tada:2023rgp}. In obtaining our cubic and quartic Hamiltonians presented above, we have discarded the total time derivative terms. These total time derivative terms are specifically presented in 
\cite{Firouzjahi:2023aum} which are shown to have the  form $\frac{d}{dt} (f(\tau) \pi^3) $ and $\frac{d}{dt} (g(\tau) \pi^4) $ in which $f(\tau)$ and $g(\tau)$ are functions of background parameters like  $\epsilon$. It turns out that when we calculate the correlations in power spectrum $\langle \pi  \pi \rangle $ or bispectrum $\langle \pi  \pi  \pi \rangle $, involving only the fields and not their derivatives $\dot \pi$, then  these total time derivative terms do not contribute. This is because the fields commute at the final point (end of inflation). However, this would not be the case if the total time derivative terms involve terms with derivative of fields. The roles of  total time derivative terms are more closely investigated recently in \cite{Braglia:2024zsl}. 

\section{Loop Corrections in Bispectrum}

We are interested in loop corrections in bispectrum $\langle \calR_{\bfp_1} \calR_{\bfp_2} \calR_{\bfp_3} \rangle$ in which we assume all three modes $\bfp_i$ are on the CMB scales. For simplicity, we assume $p_1 \simeq p_2 \simeq p_3$ and they are considered soft momenta compared to mode $\bfq$ which leaves the horizon during the USR phase:  $p_i \ll q$. We neglect the tree-level non-Gaussianity in $\langle \calR_{\bfp_1} \calR_{\bfp_2} \calR_{\bfp_3} \rangle$ 
which are induced from the gravitational back-reactions and are negligible in the limit of slow-roll approximation \cite{Maldacena:2002vr}. 

To calculate the loop corrections in bispectrum we employ the perturbative in-in formalism ~\cite{Weinberg:2005vy} in which the expectation value of the operator $\hat {O}[\tau_0]$ at the end of inflation $\tau_0$ is given by, 
 \begin{equation}
 \label{Dyson}
 \langle \hat O(\tau_0) \rangle = \Big \langle \Big[ \bar {\mathrm{T}} \exp \Big( i \int_{-\infty}^{\tau_0} d \tau' H_{\mathrm{in}} (\tau') \Big) \Big] \,  \hat O(\tau_0)  \, \Big[ \mathrm{T} \exp \Big( -i \int_{-\infty}^{\tau_0} d \tau' H_{\mathrm{in}} (\tau') \Big) \Big]
 \Big \rangle \, ,
 \end{equation}
 in which $\mathrm{T}$ and $\bar {\mathrm{T}}$ represent the time ordering and anti-time ordering respectively while $H_{\mathrm{in}}(t)$ is the interaction Hamiltonian. In our case with $ \hat O(\tau_0)= \calR_{\bfp_1}(\tau_0) \calR_{\bfp_2}(\tau_0) \calR_{\bfp_3}(\tau_0)$, we have the contributions from the cubic, quartic and quintic  interactions so  $H_{\mathrm{in}} = {\bf H}_3 + {\bf H}_4 + {\bf H}_5$.
 
As mentioned before, there are three Feynman diagrams  at the one-loop level as shown in Fig. \ref{Feynman-fig}. Since the interaction vertices in the left panel of this figure all  involve ${\bf H_3}$, then this diagram  requires three factors of ${\bf H_3}$ inside the in-in integrals. Correspondingly, one would deal with a threefold time integrals over $\tau_1, \tau_2, \tau_3$ which make the analysis complicated, see  \cite{Agui-Salcedo:2023wlq} for recent progress in dealing with the nested time integrals.  On the other hand, the middle panel in Fig. \ref{Feynman-fig}  involves one factor of ${\bf H_3}$ and one factor of  ${\bf H_4}$ so it contains a  twofold time integrals over $\tau_1$ and $\tau_2$ and the analysis for this diagram would be easier. 
Finally, the diagram in the right panel of this figure contains only one vertex 
of ${\bf H_5}$ so it would involve only a single time integral. However, for this diagram one has to calculate  total ${\bf H_5}$. In this work, to obtain a first estimation of the one-loop corrections in bispectrum, we consider the diagram in the middle panel of Fig. \ref{Feynman-fig} in which the interaction Hamiltonians are already known and the in-in integrals are easier compared to the diagram in the left panel. 
Of course, to find the total loop corrections one has to calculate the contribution of all diagrams in Fig. \ref{Feynman-fig}.

To calculate the loop corrections, we need to expand the in-in formula Eq. (\ref{Dyson}) to desired order in powers of $H_{\mathrm{in}}$. For the diagram in middle panel of Fig. \ref{Feynman-fig}, we have to expand to second order in $H_{\mathrm{in}}$. It turns out to be more convenient to employ the Weinberg method of commutator series of master equation (\ref{Dyson}) which, to second order in interaction,   yields \cite{Weinberg:2005vy},
\ba
\label{Weinberg}
 \langle \hat O (\tau_0) \rangle = i^2 \int_{-\infty}^{\tau_0} d\tau_2 \int_{-\infty}^{\tau_2} d \tau_1 \Big \langle \Big[ H_{\mathrm{in}}(\tau_1) , \big[ H_{\mathrm{in}}(\tau_2), \hat O (\tau_0) \big] \,  \Big] \Big \rangle \, ,
\ea
in which, excluding the quintic Hamiltonian,   $H_{\mathrm{in}} = {\bf H}_3 + {\bf H}_4$. To proceed further, note that
since $\calR$ is a free Gaussian field, the expectation value  $\langle \calR^n \rangle$ vanishes for odd values of $n$. Correspondingly,  the above formula is cast into 
\ba
\label{Weinberg2}
 \langle \calR_{\bfp_1}(\tau_0) \calR_{\bfp_2}(\tau_0) \calR_{\bfp_3}(\tau_0) \rangle  =  {\cal B} + {\cal C} \, ,
\ea 
where the two contributions ${\cal B}$ and ${\cal C}$ are given as follows,
\ba
\label{B-def}
{\cal B}  &\equiv&  \int_{-\infty}^{\tau_0} d\tau_2 \int_{-\infty}^{\tau_2} d \tau_1 
 \Big \langle \big[ {\bf H_4} (\tau_1) , \big[\hat O (\tau_0),  {\bf H_3} (\tau_2) \big]   \big] \Big \rangle  \nonumber\\
 &=& 2\int_{-\infty}^{\tau_0} d\tau_2 \int_{-\infty}^{\tau_2} d \tau_1
 \mathrm{Re} \Big[ \big \langle {\bf H_4} (\tau_1)  \hat O (\tau_0)  {\bf H_3} (\tau_2) \big \rangle 
 - \big \langle {\bf H_4} (\tau_1) {\bf H_3} (\tau_2)  \hat O (\tau_0)   \big \rangle 
 \Big] \, ,
\ea
and
\ba
\label{C-def}
{\cal C}  &\equiv&  \int_{-\infty}^{\tau_0} d\tau_2 \int_{-\infty}^{\tau_2} d \tau_1 
 \Big \langle \big[ {\bf H_3} (\tau_1) , \big[\hat O (\tau_0),  {\bf H_4} (\tau_2) \big]   \big] \Big \rangle  \nonumber\\
 &=& 2\int_{-\infty}^{\tau_0} d\tau_2 \int_{-\infty}^{\tau_2} d \tau_1
 \mathrm{Re} \Big[ \big \langle {\bf H_3} (\tau_1)  \hat O (\tau_0)  {\bf H_4} (\tau_2) \big \rangle 
 - \big \langle {\bf H_3} (\tau_1) {\bf H_4} (\tau_2)  \hat O (\tau_0)   \big \rangle 
 \Big] \, ,
\ea
with $\hat O (\tau_0) \equiv \calR_{\bfp_1}(\tau_0) \calR_{\bfp_2}(\tau_0) \calR_{\bfp_3}(\tau_0)$. 
Note that the contributions 
$ \big[ {\bf H_4} (\tau_1) , \big[ {\bf H_4} (\tau_2), \hat O (\tau_0) \big]   \big]$
and $\big[ {\bf H_3} (\tau_1) , \big[ {\bf H_3} (\tau_2), \hat O (\tau_0) \big]   \big]$ which lead to odd powers of $\calR$ vanish as just mentioned above. Also we note that the time integral is nested with
$-\infty< \tau_1 \leq \tau_2 \leq \tau_0$.
Finally, as noted before, the relation between $\pi$ and $\calR$ can be considered linear inside the in-in integral (see discussions after Eq. (\ref{pi-R})) so
we replace $\pi$ in ${\bf H_3}$ and ${\bf H_4}$ by  $\calR$.

Looking at the form of ${\bf H_3}$ and ${\bf H_4}$ we observe that both are divided to two different forms: either having   time derivatives like $\pi'^2$ or having gradient like $\partial^2 \pi$. Therefore, to keep track of their  contributions inside the in-in integrals we decompose them as follows,
\ba
\label{H3-AB}
{\bf H_3} =  A_3 (\tau) \int d^3 \bfx \, \calR \calR'^2 + 
B_3(\tau)  \int d^3 \bfx \, \calR^2 \partial^2  \calR \, ,
\ea
and
\ba
\label{H4-AB}
{\bf H_4} =  A_4 (\tau) \int d^3 \bfx \, \calR^2 \calR'^2 + 
B_4(\tau)  \int d^3 \bfx \, \calR^2  (\partial  \calR)^2 \, ,
\ea
in which the time-dependent coefficients $A_i$ and $B_i$ are defined via,
\ba
A_3(\tau)  = 2 B_3( \tau)  \equiv M_P^2  \eta \epsilon a^2 \, , 
\ea
and
\ba
A_4(\tau)    \equiv  \frac{1}{2} M_P^2 \eta^2 \epsilon a^2 \Big(  1- \frac{h}{\eta^2} \delta (\tau -\tau_e)  \tau_e \Big) \, , \quad \quad 
B_4(\tau)    \equiv  \frac{1}{2} M_P^2 \eta^2 \epsilon a^2 \Big(  1+ \frac{h}{\eta^2} \delta (\tau -\tau_e)  \tau_e \Big) \, .
\ea
Note that the term $\delta (\tau -\tau_e) $  above comes from the term $\eta'$ in 
${\bf H_4}$ as given in  Eq. (\ref{eta-jump}).

Depending on which terms from  the above two categories in ${\bf H_3}$ and ${\bf H_4}$ are contracted with each other, we will have four different contributions in 
each component ${\cal B}$  and ${\cal C}$ of bispectrum 
as follows:
\ba
{\cal B}
= {\cal B}_{A_4 A_3}+ {\cal B}_{A_4 B_3}+ {\cal B}_{B_4 A_3} 
+ {\cal B}_{B_4 B_3} \, ,
\ea
and
\ba
{\cal C}
= {\cal C}_{A_3 A_4}+ {\cal C}_{A_3 B_4}+ {\cal C}_{B_3 A_4} 
+ {\cal C}_{B_3 B_4} \, . 
\ea
For example, for ${\cal B}_{A_4 A_3}$ we have
\ba
\label{B-AA}
{\cal B}_{A_4 A_3} = \int_{-\infty}^{\tau_0} d \tau_2 \int_{-\infty}^{\tau_2} d \tau_1 A_4(\tau_1) A_3(\tau_2)  \int d^3 \bfx \int d^3  {\bf y}
\Big \langle  \Big[ \calR^2 \calR'^2 (\bfx, \tau_1) ,  \big[ \hat O (\tau_0)  ,  \calR \calR'^2 (\bf{y}, \tau_2) \big] \Big]
 \Big \rangle \, .
\ea
Going to the Fourier space, this is cast into 
\ba
\label{B-AA}
{\cal B}_{A_4 A_3}  =  \int_{-\infty}^{\tau_0} d \tau_2 \int_{-\infty}^{\tau_2} d \tau_1  A_4(\tau_1) A_3(\tau_2) 
\Big[ \prod_i^4 \int \frac{d^3 \bfq_i}{ (2 \pi)^\frac{3}{2}} (2 \pi)^3 \delta^3( \sum_i \bfq_i)\Big] 
\Big[\prod_j^3 \int \frac{d^3 \bfk_i}{(2 \pi)^\frac{3}{2}} (2 \pi)^3 \delta^3( \sum_i \bfk_i)
\Big]  \hspace{-1cm} \nonumber\\
 \times  \bigg \langle   \bigg[   \big( \hat\calR_{\bfq_1}   \hat\calR_{\bfq_2} \hat\calR'_{\bfq_3} \hat\calR'_{\bfq_4} \big) (\tau_1), \Big[
 \big( \hat\calR_{\bfp_1} \hat\calR_{\bfp_2} \hat\calR_{\bfp_3} \big) (\tau_0)  , \big( \hat\calR_{\bfk_1}  \hat\calR'_{\bfk_2}   \hat\calR'_{\bfk_3} \big) (\tau_2)  \Big]  \bigg]  \bigg \rangle \, .
\ea

We present the details of the analysis for the in-in integrals in the Appendix \ref{appendix}.  After a tedious and long calculations, we obtain
\ba
\label{B-result}
{\cal B}'  = -16 P_\calR(p_1) P_\calR(p_2)  \int \frac{d^3 \bf q}{( 2 \pi)^3}  
\int_{-\infty}^{\tau_0}   d \tau_2   \int_{-\infty}^{\tau_2} d \tau_1
\bigg\{\mathrm{Im} \big[ \calR_{p}(\tau_0)  \calR^*_{p}(\tau_2)  \big] 
\mathrm{Im}\big[ I_4 (q, \tau_1) I_3(q, \tau_2) \big] \hspace{.5cm}  \\
+2 \mathrm{Im} \big[ \calR_{p}(\tau_0)  \calR^{'*}_{p}(\tau_2)  \big] 
\mathrm{Im} \big[ I_4 (q, \tau_1) \calR^*_q(\tau_2) \calR^{'*}_q(\tau_2)
\big]  A_3(\tau_2) \bigg\} +  2\, \mathrm{c. p.}
\nonumber
\ea
in which the prime over ${\cal B}$ and  similar expressions below  means that 
we have pulled out  the factor $(2 \pi)^3 \delta^3 (\bfp_1+ \bfp_2 + \bfp_3)$ and 
$ \mathrm{c. p.}$ means cyclic permutations over $p_i$. 
In addition,  the quantities $I_4(q, \tau)$ and $I_3(q, \tau)$ are defined as follows:
\ba
I_4(q, \tau) \equiv  
 \Big[ A_4 (\tau) \calR_q'(\tau)^2  + q^2 B_4(\tau)  \calR_q(\tau)^2 \Big]  \, ,
\ea
and
\ba
I_3(q, \tau) \equiv  \Big[ A_3 (\tau) \calR_q^{'*}(\tau)^2  -2 q^2 B_3(\tau)  \calR_q^*(\tau)^2 \Big] \, .
\ea 
Similarly, for the remaining part in bispectrum we obtain (see Appendix \ref{appendix} for further details),
\ba
\label{C-result}
{\cal C}'  = -32 P_\calR(p_1) P_\calR(p_2)    \int \frac{d^3 \bf q}{( 2 \pi)^3}  
\int_{-\infty}^{\tau_0}   d \tau_2   \int_{-\infty}^{\tau_2} d \tau_1  
\bigg\{\mathrm{Im} \big[ \calR_{p}(\tau_0)  \calR^*_{p}(\tau_2)  \big] 
\mathrm{Im}\big[ I_4^* (q, \tau_2) I_3^*(q, \tau_1) \big]  \hspace{0.5cm} \\
+2 \mathrm{Im} \big[ \calR_{p}(\tau_0)  \calR^{'*}_{p}(\tau_2)  \big] 
\mathrm{Im} \big[ I_3^* (q, \tau_1) \calR^*_q(\tau_2) \calR^{'*}_q(\tau_2)
\big]  A_4(\tau_2) \bigg\} +  2\, \mathrm{c. p.} \, .
\nonumber
\ea

In the expressions of ${\cal B}'$ and ${\cal C}'$ the  imaginary factors
$ \mathrm{Im} \big[ \calR_{p}(\tau_0)  \calR^{*}_{p}(\tau_2)  \big]$ and 
$ \mathrm{Im} \big[ \calR_{p}(\tau_0)  \calR^{'*}_{p}(\tau_2)  \big]$
are given by \cite{Firouzjahi:2023aum}
\begin{equation}
 \mathrm{Im}  \big[  \calR_p(\tau_0)  \calR_p^*(\tau)   \big]  = \frac{-H^2 \tau_s^6}{12 M_P^2 \epsilon_i  h \tau_e^3 \tau^3 } \big( h \tau_e^3 + (6- h) \tau^3
 \big) \, ,
\end{equation}
and
\begin{equation}
 \mathrm{Im}  \big[  \calR_p(\tau_0)  \calR'^*_p(\tau)   \big]  = \frac{H^2 \tau_s^6}{4 M_P^2 \epsilon_i   \tau^4 }  \, .
\end{equation}
Note that the above imaginary components are independent 
of $p$. This played important roles in simplifying the analysis in obtaining 
the results for ${\cal B}$ and ${\cal C}$ in Eqs. (\ref{B-result}) and (\ref{C-result}).  

Looking at the momentum dependence of ${\cal B}'$ and ${\cal C}'$ we observe that the loop correction in bispectrum  has the local shape in its dependence on $p_i$. This is an interesting result, as it is expected usually 
that the local shape bispectrum to occur  in multiple fields scenarios \cite{Wands:2007bd}. In this view the loop correction in the current single field model  provides a counter example for this general expectation. 

Now defining the $f_{NL}$ parameter via, 
\ba
\label{fNL-def}
\big \langle \calR_{\bfp_1}(\tau_0) \calR_{\bfp_2}(\tau_0) \calR_{\bfp_3}(\tau_0) \big \rangle \equiv \frac{6}{5} f_{NL} \Big( P_\calR(p_1, \tau_0)  P_\calR(p_2, \tau_0) + 2\, \mathrm{c. p.} \Big) \, ,
\ea
from our expressions for ${\cal B}' $ and ${\cal C}'$ given above, we obtain
\ba
\label{fNL-loop}
f_{\mathrm{NL}}^{\mathrm{loop}} =-\frac{40}{3} 
 \int \frac{d^3 \bf q}{( 2 \pi)^3} 
\int_{-\infty}^{\tau_0}   d \tau_2   \int_{-\infty}^{\tau_2} d \tau_1 \, {\cal F}_q(\tau_2, \tau_1) \, ,
\ea
in which the kernel function ${\cal F}_q(\tau_2, \tau_1)$ is defined via 
\ba
\label{kernel}
&&{\cal F}_q(\tau_2, \tau_1) \equiv  
\mathrm{Im} \big[ \calR_{p}(\tau_0)  \calR^*_{p}(\tau_2)  \big] 
\bigg\{  \mathrm{Im}\big[ I_4 (q, \tau_1) I_3(q, \tau_2) \big] + 2\mathrm{Im}\big[ I_4^* (q, \tau_2) I_3^*(q, \tau_1) \big] \bigg\} \\
&&  +2 \mathrm{Im} \big[ \calR_{p}(\tau_0)  \calR^{'*}_{p}(\tau_2)  \big]
\bigg\{ A_3(\tau_2) \mathrm{Im} \big[ I_4 (q, \tau_1) \calR^*_q(\tau_2) \calR^{'*}_q(\tau_2)
\big]   + 2 A_4(\tau_2) \mathrm{Im} \big[ I_3^* (q, \tau_1) \calR^*_q(\tau_2) \calR^{'*}_q(\tau_2)
\big]  \bigg\}  \nonumber
\ea

To estimate $f_{\mathrm{NL}}^{\mathrm{loop}}$ we consider the contributions of the USR mode with $q_s \leq q \leq q_e$. In addition, we only count the modes which become superhorizon during the USR phase with $-q \tau <1$. In performing the integral over time, we consider the contributions from the interval $\tau_s \leq \tau \leq \tau_e$, including the contribution from the local source term $\delta (\tau- \tau_e)$. However, as shown in  \cite{Firouzjahi:2023aum}, the contributions from the final stage of inflation $\tau_e < \tau < \tau_0$  in the time integrals are negligible. This is because in our approximation of a sharp transition, the mode function quickly approaches its final attractor phase and their contributions in the time integral is not significant.

Performing the in-in integrals\footnote{We use the Maple computational  software to calculate the integrals semi-analytically.}, we obtain the following result for $f_{\mathrm{NL}}^{\mathrm{loop}}$,
\ba
\label{fNL-loop}
f_{\mathrm{NL}}^{\mathrm{loop}} = 45 f(h) \Delta N  \calP_{\mathrm{CMB}}  e^{6 \Delta N}  \, ,
\ea
in which $\calP_{\mathrm{CMB}} \simeq 2 \times 10^{-9}$ is the power spectrum on the CMB scales and   the function $f(h)$ is given by,
\ba
\label{fh}
f(h) \simeq \frac{3 - 5.9 h+ 0.36 h^2 }{h} \, .
\ea
In particular, we note that for $\Delta N=0$, the loop correction in bispectrum vanishes. This is expected since when $\Delta N=0$ there is no 
intermediate USR phase so all interactions disappear and we obtain the tree level results that $f_{NL} \rightarrow 0$ in the slow-roll limit.

The expression (\ref{fNL-loop}) for the loop correction in bispectrum has the same structure as the loop corrections in power spectrum with the factor $ \Delta N  \calP_{\mathrm{CMB}}  e^{6 \Delta N}$ appearing  in both analysis. The non-linear dependence on the sharpness parameter $h$ in $f(h)$
is similar but with different numerical factors.  More specifically, the fractional loop correction in power spectrum is calculated in  \cite{Firouzjahi:2023aum} to be 
\begin{equation}
\label{fraction}
\Delta^{\mathrm{loop}}\equiv 
\frac{\Delta \calP^{\mathrm{loop}}}{\calP_{\mathrm{CMB}}} = \frac{6 \Delta N}{h}
   (  h^2  + 24 h + 180)  e^{6 \Delta N} \calP_{\mathrm{CMB}} \, .
\end{equation}
Combining this result with our loop correction in bispectrum, we can eliminate the common factor $\Delta N  \calP_{\mathrm{CMB}}  e^{6 \Delta N}$ and obtain the following relation between the loop corrections in power spectrum and bispectrum:
\ba
\label{power-fNL}
f_{\mathrm{NL}}^{\mathrm{loop}} \simeq 
 \frac{ 15( 0.36 h^2- 5.9 h + 3) } {2(h^2  + 24 h + 180)}  \Delta^{\mathrm{loop}} \, .
\ea
For the perturbative analysis to be under control we need that $| \Delta^{\mathrm{loop}}| <1$. This in turn imposes a theoretical bound on $f_{\mathrm{NL}}^{\mathrm{loop}}$ and the sharpness parameter $h$. We comment that in \cite{Kristiano:2021urj}
the authors used the one-loop corrections in power spectrum to put a bound on 
the tree-level $f_{NL}$ in standard single field models of inflation.

\begin{figure}[t]
\vspace{-1 cm}
	\centering
	\includegraphics[ width=0.49\linewidth]{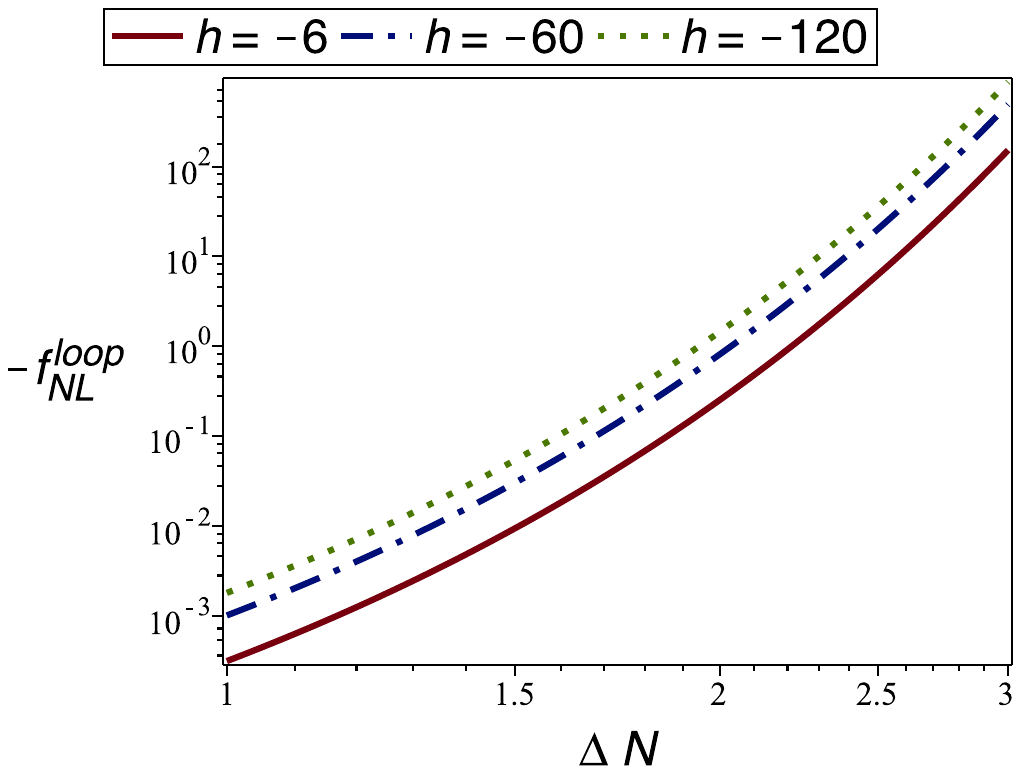}
	\includegraphics[ width=0.49 \linewidth]{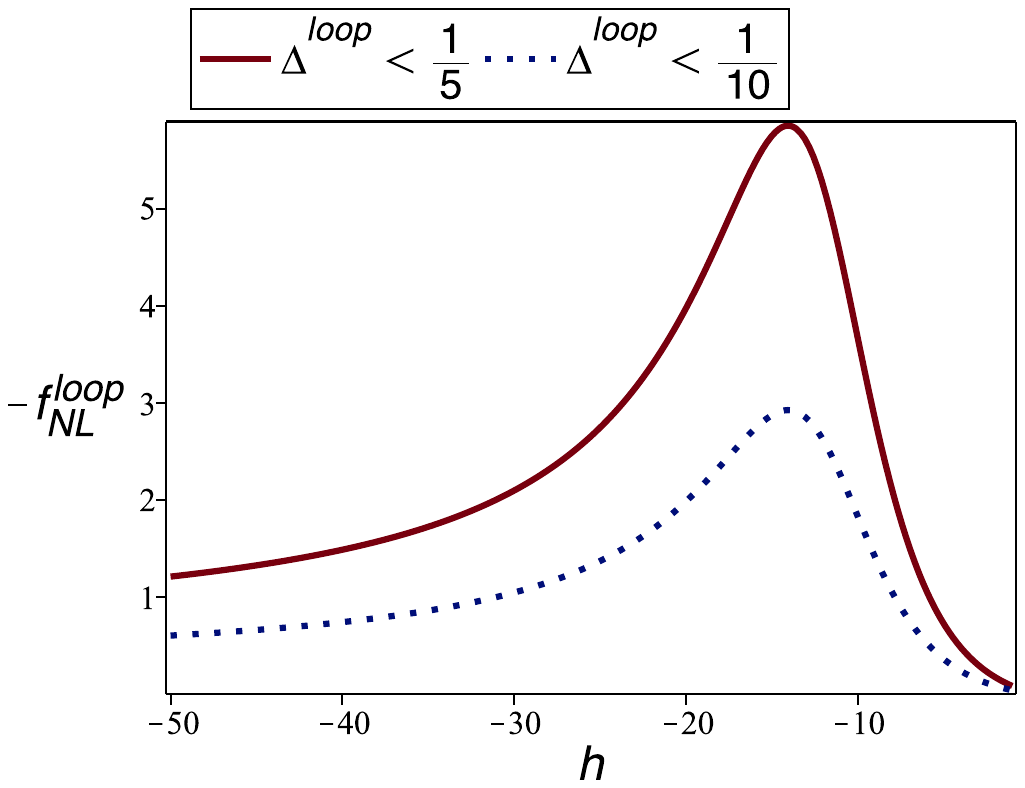}
	\vspace{-4.5 cm}
	\caption{ Left: the plot of $-f_{\mathrm{NL}}^{\mathrm{loop}}$ vs. $\Delta N$ 
	for different values of $h$. From bottom to top:  $h=-6$, $h=-60$  and $h=-120$.  Right: the upper bound on   $-f_{\mathrm{NL}}^{\mathrm{loop}}$  from Eq. (\ref{power-fNL}) for different upper bounds on  the fractional loop corrections in power spectrum $\Delta^{\mathrm{loop}} $:  $| \Delta^{\mathrm{loop}} |<\frac{1}{5}$ (solid) and  $| \Delta^{\mathrm{loop}}| <\frac{1}{10}$ (dot). 
}
\label{fNL-fig}
\end{figure}

In the left panel of Fig. \ref{fNL-fig} we have plotted $f_{\mathrm{NL}}^{\mathrm{loop}}$ for various values of $h$ as a function of the duration of USR phase $\Delta N$. We see the dependence on $h$ is moderate, but $|f_{NL}|$ increases linearly for $|h| 
\gg 1$.   Furthermore,   $f_{\mathrm{NL}}^{\mathrm{loop}}$ increases quickly beyond the observational bound  \cite{Planck:2018jri} $|f_{NL} |\lesssim 10$ for $N \gtrsim 2.5$. The conclusion is that to satisfy the observational bound on $f_{NL}$ the duration of USR phase should be limited to $\Delta N \lesssim 2.5$ for sharp transitions with $| h| >1$. In addition, our analysis suggests that $f_{\mathrm{NL}}^{\mathrm{loop}}<0$. This may relax the PBHs formation 
with a negative local-type non-Gaussianity \cite{Byrnes:2012yx, Young:2015kda, Atal:2018neu, Firouzjahi:2023xke, Namjoo:2024ufv}. On the other hand,  in the right panel of Fig. \ref{fNL-fig} we have presented the bound on $| f_{\mathrm{NL}}^{\mathrm{loop}}|$ for different 
theoretical upper bounds on $\Delta^{\mathrm{loop}} $ based on Eq. (\ref{power-fNL}).   For moderate large value of $|h|$ the theoretical bound (\ref{power-fNL}) is similar to the observational bound $|f_{NL} |\lesssim 10$.  However, for extreme sharp transition with $h \rightarrow -\infty$, the theoretical bound is stronger, requiring $| f_{\mathrm{NL}}^{\mathrm{loop}}| \lesssim 1$. Curiously, for relatively mild transition with $ 1\lesssim |h| \lesssim 6$ the upper bound on $| f_{\mathrm{NL}}^{\mathrm{loop}}| $ becomes strong as well while we may not trust our analysis for $|h|\ll1$  where the assumption of a sharp transition is violated.

Large values of $f_{\mathrm{NL}}^{\mathrm{loop}}$ can be constrained by their implications for gravitational anisotropies and dark matter isocurvature  perturbations \cite{Young:2015kda, Bartolo:2019zvb}. More specifically, consider a constant and scale invariant $f_{NL}$. 
The modulation of large CMB scale perturbations on PBHs induces isocurvature perturbations of amplitude  \cite{Young:2015kda}
$\calP_{\mathrm{iso}} \sim f_{NL}^2 f_{\mathrm{PBH}}^2 \calP_{\mathrm{CMB}}$
in which $f_{\mathrm{PBH}}$ is the fraction of PBHs in dark matter energy density. 
From the Planck constraints on isocurvature perturbations \cite{Planck:2018jri} one typically requires that  $ | f_{NL} f_{\mathrm{PBH}} | \lesssim 10^{-2}$. An immediate conclusion is that for large value of $f_{NL}$, the PBHs can not furnish a significant fraction of the dark matter energy density. In addition, a large value of $f_{NL}$ induces anisotropies and non-Gaussianities in stochastic GWs spectra generated from the second order scalar perturbations which can be constrained as well \cite{Bartolo:2019zvb}.

\section{Summary and Discussions } 
\label{Summary}

Motivated by the analysis of \cite{Kristiano:2022maq, Kristiano:2023scm}, in this work  we have studied the one-loop corrections in bispectrum for the CMB scale perturbations induced from the small scale modes which undergo an intermediate phase of USR inflation. To simplify the analysis we have assumed a sharp transition to the final attractor phase. These setups with an intermediate USR phase have been employed to amplify the small scales perturbations   for the PBHs formation during inflation. 

There are three distinct one-particle irreducible Feynman diagrams for the one-loop corrections in bispectrum as  depicted in Fig. \ref{Feynman-fig}.  For simplicity,  we have considered the Feynman diagram with two exchange vertices, one from the cubic interaction and  one from the  quartic interactions.  We have employed the formalism of EFT of inflation to calculate the cubic and quartic Hamiltonians. After performing the in-in analysis, we have shown that the bispectrum has the local shape with the magnitude  $f_{\mathrm{NL}}^{\mathrm{loop}} \propto f(h)  \calP_{\mathrm{CMB}}  e^{6 \Delta N}$. The dependence on $\calP_{\mathrm{CMB}}  e^{6 \Delta N}$ is the same as in the loop corrections for power spectrum $\Delta^{\mathrm{loop}}$. In addition, the function $f(h)$ increases linearly with the sharpness parameter $h$ for $|h|\gg1$ which is similar to the one-loop corrections in power spectrum  \cite{Firouzjahi:2023aum}. Our analysis suggests that $f_{\mathrm{NL}}^{\mathrm{loop}}$ associated to this Feynman diagram is universally negative. This can help the bounds on the PBHs formation. 

We have found  that for sharp transitions 
the loop corrections in bispectrum can quickly violate the observational bounds on $f_{NL}$ requiring $| f_{NL}| \lesssim 10$ 
on CMB scales  \cite{Planck:2018jri}. This imposes an upper bound on the duration of USR phase, requiring $\Delta N \lesssim 2.5$ or so. On the other hand, one can also put a theoretical bound on $f_{\mathrm{NL}}^{\mathrm{loop}}$ by requiting that the loop corrections in power spectrum to be under perturbative control with 
$|\Delta^{\mathrm{loop}} |<1$. 
 For moderate large values of $|h|$ this theoretical bound is similar to the observational bound $| f_{NL}| \lesssim 10$ but for extreme sharp transition with $|h| \gg 10$, one obtains the strong theoretical bound $| f_{\mathrm{NL}}^{\mathrm{loop}}| \lesssim 1$. 

For a complete understanding of the overall sign and magnitude of $ f_{\mathrm{NL}}^{\mathrm{loop}}$ we  need to calculate the contribution 
from the remaining two diagrams of  Fig. \ref{Feynman-fig} as well. The analysis from  \cite{Firouzjahi:2023aum}  for the case of power spectrum 
suggests that the three diagrams in Fig. \ref{Feynman-fig} can have similar magnitudes. Having said this, one  expects that the contributions from the  three diagrams show different properties as well. For example,   the interaction Hamiltonian ${\bf H_4}$ contains $\eta' \propto h$ while the interaction  ${\bf H_3}$ does not. As such, the dependence on the sharpness parameter $h$ would be somewhat different for the one-loop corrections induced by these two interactions. We leave it for future work to calculate the one-loop corrections from all three diagrams  in Fig. \ref{Feynman-fig}. As part of this analysis, we need  the quintic interaction Hamiltonian ${\bf H_5}$ which is not calculated previously. 

There are a number of directions in which the current study can be extended.  In this work we only considered the effects of USR modes inside the loop integrals but did not take into account the UV scales corresponding to the modes which are deep inside the horizon during the USR phase. This brings the important question of renormalization and regularization as in any QFT analysis. However, we believe the result given in Eq. (\ref{fNL-loop}) provides a useful and reliable estimation of the one-loop corrections in bispectrum.  We leave the question of regularization and renormalization for future studies. Another question of interest is to look for the one-loop corrections in trispectrum. There are various Feynman diagrams corresponding for the one-loop corrections in trispectrum, including the contributions from the sixth order Hamiltonian ${\bf H_6}$.  It would be interesting to examine if the one-loop correction in the trispectrum has the same structure as what is obtained for power spectrum and bispectrum.

\vspace{1.2cm}
   
 {\bf Acknowledgments:}  We thank Antonio Riotto, Mohammad Hossein Namjoo, Jason Kristiano, Amin Nassiri-Rad and  Kosar Asadi    for useful discussions and correspondences.   This work is supported by INSF  of Iran under the grant  number 4025208. 
  
\vspace{1.5cm}
  
\appendix
\section{In-In Integrals }  
 \label{appendix}
In this Appendix we present the in-in analysis in more details. 

As mentioned in the main text, there are two different contributions for the one-loop corrections in  bispectrum  associated to the diagram in the middle panel of Fig. \ref{Feynman-fig} as follows, 
\ba
\label{Weinberg2b}
 \langle \calR_{\bfp_1}(\tau_0) \calR_{\bfp_2}(\tau_0) \calR_{\bfp_3}(\tau_0) \rangle  =  {\cal B} + {\cal C} \, ,
\ea 
with 
\ba
\label{B-defb}
{\cal B}  &\equiv&  \int_{-\infty}^{\tau_0} d\tau_2 \int_{-\infty}^{\tau_2} d \tau_1 
 \Big \langle \big[ {\bf H_4} (\tau_1) , \big[\hat O (\tau_0),  {\bf H_3} (\tau_2) \big]   \big] \Big \rangle  \nonumber\\
 &=& 2\int_{-\infty}^{\tau_0} d\tau_2 \int_{-\infty}^{\tau_2} d \tau_1
 \mathrm{Re} \Big[ \big \langle {\bf H_4} (\tau_1)  \hat O (\tau_0)  {\bf H_3} (\tau_2) \big \rangle 
 - \big \langle {\bf H_4} (\tau_1) {\bf H_3} (\tau_2)  \hat O (\tau_0)   \big \rangle 
 \Big] \, ,
\ea
and
\ba
\label{C-defb}
{\cal C}  &\equiv&  \int_{-\infty}^{\tau_0} d\tau_2 \int_{-\infty}^{\tau_2} d \tau_1 
 \Big \langle \big[ {\bf H_3} (\tau_1) , \big[\hat O (\tau_0),  {\bf H_4} (\tau_2) \big]   \big] \Big \rangle  \nonumber\\
 &=& 2\int_{-\infty}^{\tau_0} d\tau_2 \int_{-\infty}^{\tau_2} d \tau_1
 \mathrm{Re} \Big[ \big \langle {\bf H_3} (\tau_1)  \hat O (\tau_0)  {\bf H_4} (\tau_2) \big \rangle 
 - \big \langle {\bf H_3} (\tau_1) {\bf H_4} (\tau_2)  \hat O (\tau_0)   \big \rangle 
 \Big] \, ,
\ea
with $\hat O (\tau_0) \equiv \calR_{\bfp_1}(\tau_0) \calR_{\bfp_2}(\tau_0) \calR_{\bfp_3}(\tau_0)$. 
Note that  ${\cal B}$ and ${\cal C}$ are mirror to each other concerning the 
positions of ${\bf H_3}$ and ${\bf H_4}$. 

To proceed further, it is very convenient to decompose the expectation values in terms of sub-component Wick contractions. For example, consider the expression 
$\big \langle {\bf H_4} (\tau_1)  \hat O (\tau_0)  {\bf H_3} (\tau_2) \big \rangle$ in the second line for ${\cal B}$. It has three types of Wick contractions: 
$\contraction[1ex]{}{bc}{eeem}
({\bf H_4}(\tau_1){\bf H_3}(\tau_2)$,  $\contraction[1ex]{}{bc}{eeem}
({\bf H_4}(\tau_1) \hat O (\tau_0)$ 
and  $\contraction[1ex]{}{bc}{eem}
(\hat O (\tau_0) {\bf H_3}(\tau_2)$.
Let us define
\ba
\contraction[1ex]{}{bc}{eeem}
({\bf H_4}(\tau_1){\bf H_3}(\tau_2) \equiv \hat h(\tau_1, \tau_2)  \, , \qquad 
\contraction[1ex]{}{bc}{eeem}
({\bf H_4}(\tau_1) \hat O (\tau_0)\equiv \hat f(\tau_1)   \, , \qquad 
\contraction[1ex]{}{bc}{eem}
(\hat O (\tau_0) {\bf H_3}(\tau_2) \equiv \hat g(\tau_2)  \, .
\ea
Then, considering both terms in the second line of Eq. (\ref{B-defb}), 
the expression for ${\cal B}$ is simplified to
\ba
\label{B-eq}
{\cal B}  = -4   \int_{-\infty}^{\tau_0} d\tau_2 \int_{-\infty}^{\tau_2} d \tau_1 
 \mathrm{Im} [\hat g(\tau_2)] \,  \mathrm{Im}\big[ \hat f(\tau_1) \hat h(\tau_1 , \tau_2)
 \big]  \, .
\ea
Our job would be to calculate the functions $\hat g(\tau_2), \hat f(\tau_1)$ and  $\hat h(\tau_1 , \tau_2)$ for each case and plug them into Eq. (\ref{B-eq}).

Depending on the relative positions of $A_i$ and $B_i$ terms in   ${\bf H_3}$ 
and ${\bf H_4}$ interactions,  ${\cal B}$ and
${\cal C}$ are decomposed in four different contributions, 
\ba
{\cal B} = {\cal B}_{A_4 A_3}+ {\cal B}_{A_4 B_3}+ {\cal B}_{B_4 A_3} 
+ {\cal B}_{B_4 B_3} \, ,
\ea
and
\ba
{\cal C} = {\cal C}_{A_3 A_4}+ {\cal C}_{A_3 B_4}+ {\cal C}_{B_3 A_4} 
+ {\cal C}_{B_3 B_4} \, . 
\ea
For example, for  ${\cal B}_{A_4 A_3}$ we have  
\ba
\label{B-AA}
{\cal B}_{A_4 A_3}  =  \int_{-\infty}^{\tau_0} d \tau_2 \int_{-\infty}^{\tau_2} d \tau_1  A_4(\tau_1) A_3(\tau_2) 
\Big[ \prod_i^4 \int \frac{d^3 \bfq_i}{ (2 \pi)^\frac{3}{2}} (2 \pi)^3 \delta^3( \sum_i \bfq_i)\Big] 
\Big[\prod_j^3 \int \frac{d^3 \bfk_i}{(2 \pi)^\frac{3}{2}} (2 \pi)^3 \delta^3( \sum_i \bfk_i)
\Big]  \hspace{-1cm} \nonumber\\
 \times  \bigg \langle   \bigg[   \big( \hat\calR_{\bfq_1}   \hat\calR_{\bfq_2} \hat\calR'_{\bfq_3} \hat\calR'_{\bfq_4} \big) (\tau_1), \Big[
 \big( \hat\calR_{\bfp_1} \hat\calR_{\bfp_2} \hat\calR_{\bfp_3} \big) (\tau_0)  , \big( \hat\calR_{\bfk_1}  \hat\calR'_{\bfk_2}   \hat\calR'_{\bfk_3} \big) (\tau_2)  \Big]  \bigg]  \bigg \rangle \, .
\ea
Let us concentrate on the second line above which involves the contractions
after  performing the integrals over the delta functions in the soft limit $p_i \ll q_i, k_i$.  After constructing  the functions 
$\hat g(\tau_2), \hat f(\tau_1)$ and  $\hat h(\tau_1 , \tau_2)$ for each case, we obtain 
8 different terms for these contractions as follows: 
\ba
\label{a-term}
\hspace{-1cm}
(a):
-16 \mathrm{Im} \big[ \calR_{p_3}(\tau_0)  \calR^*_{p_3}(\tau_2)  \big] 
\mathrm{Im}\big[  \calR_{p_1}^*(\tau_0)  \calR_{p_2}^*(\tau_0)   \calR_{p_1}(\tau_1) \calR_{p_2}(\tau_1)    \calR'_{q}(\tau_1)^2  {\calR_q'^*}(\tau_2)^2 \big]
+ 2\,  \mathrm{c. p.}  \nonumber
\ea
\ba
\label{b-term}
(b): 
-32 \mathrm{Im} \big[ \calR_{p_3}(\tau_0)  \calR^*_{p_3}(\tau_2)  \big] 
\mathrm{Im}\big[  \calR_{p_1}^*(\tau_0)  \calR_{p_2}^*(\tau_0)   \calR'_{p_1}(\tau_1) \calR_{p_2}(\tau_1)    \calR'_{q}(\tau_1)    \calR_{q}(\tau_1)  {\calR_q'^*}(\tau_2)^2 \big]
+ 2\,  \mathrm{c. p.}   \nonumber
\ea
\ba
\label{c-term}
(c): 
-32 \mathrm{Im} \big[ \calR_{p_3}(\tau_0)  \calR^*_{p_3}(\tau_2)  \big] 
\mathrm{Im}\big[  \calR_{p_1}^*(\tau_0)  \calR_{p_2}^*(\tau_0)   \calR_{p_1}(\tau_1) \calR'_{p_2}(\tau_1)    \calR'_{q}(\tau_1)    \calR_{q}(\tau_1)  {\calR_q'^*}(\tau_2)^2 \big]
+ 2\,  \mathrm{c. p.}  \nonumber
\ea
\ba
\hspace{-1cm}
\label{d-term}
(d): 
-16 \mathrm{Im} \big[ \calR_{p_3}(\tau_0)  \calR^*_{p_3}(\tau_2)  \big] 
\mathrm{Im}\big[  \calR_{p_1}^*(\tau_0)  \calR_{p_2}^*(\tau_0)   \calR'_{p_1}(\tau_1) \calR'_{p_2}(\tau_1)  \calR_{q}(\tau_1)^2  {\calR_q'^*}(\tau_2)^2  \big]
+ 2\,  \mathrm{c. p.}  \nonumber
\ea
\ba
\label{e-term}
(e): 
-32 \mathrm{Im} \big[ \calR_{p_3}(\tau_0)  \calR'^*_{p_3}(\tau_2)  \big] 
\mathrm{Im}\big[  \calR_{p_1}^*(\tau_0)  \calR_{p_2}^*(\tau_0)   \calR_{p_1}(\tau_1) \calR_{p_2}(\tau_1)  \calR'_{q}(\tau_1)^2  {\calR_q'^*}(\tau_2)  {\calR_q^*}(\tau_2)\big]
+ 2\,  \mathrm{c. p.} \nonumber
\ea
\ba
\label{f-term}
(f): 
-32 \mathrm{Im} \big[ \calR_{p_3}(\tau_0)  \calR'^*_{p_3}(\tau_2)  \big] 
\mathrm{Im}\big[  \calR_{p_1}^*(\tau_0)  \calR_{p_2}^*(\tau_0)   \calR'_{p_1}(\tau_1) \calR_{p_2}(\tau_1)  \calR'_{q}(\tau_1)  \calR_{q}(\tau_1) {\calR_q'^*}(\tau_2)  {\calR_q^*}(\tau_2)\big]
+ 2  \mathrm{c. p.}  \nonumber
\ea
\ba
\label{g-term}
(g): 
-32 \mathrm{Im} \big[ \calR_{p_3}(\tau_0)  \calR'^*_{p_3}(\tau_2)  \big] 
\mathrm{Im}\big[  \calR_{p_1}^*(\tau_0)  \calR_{p_2}^*(\tau_0)   \calR_{p_1}(\tau_1) \calR'_{p_2}(\tau_1)  \calR'_{q}(\tau_1)  \calR_{q}(\tau_1) {\calR_q'^*}(\tau_2)  {\calR_q^*}(\tau_2)\big]
+ 2  \mathrm{c. p.}  \nonumber
\ea
\ba
\hspace{-0.3cm}
\label{h-term}
(h): 
-32 \mathrm{Im} \big[ \calR_{p_3}(\tau_0)  \calR'^*_{p_3}(\tau_2)  \big] 
\mathrm{Im}\big[  \calR_{p_1}^*(\tau_0)  \calR_{p_2}^*(\tau_0)   \calR'_{p_1}(\tau_1) \calR'_{p_2}(\tau_1)    \calR_{q}(\tau_1)^2 {\calR_q'^*}(\tau_2)  {\calR_q^*}(\tau_2)\big]
+ 2\,  \mathrm{c. p.} \nonumber
\ea
The terms $(a), (b), (c)$ and $(d)$ share a common form of function $\hat g(\tau_2)$ while  the remaining terms $(e), (f), (g) $ and $(h)$ share a different form of function $\hat g(\tau_2)$. In the soft limit where $p_i \rightarrow 0$, the term $(a)$ from the first four terms and the term $(e)$ from the remaining four terms have the leading contributions. For example,  one can check that $(b)\sim(c)  \sim p^2 \tau \times (a)$  so these terms are suppressed compared to the (a) term. 

The above expressions for $(a)$ and $(b)$ can be further simplified noting that  \cite{Firouzjahi:2023aum}
\begin{equation}
 \mathrm{Im}  \big[  \calR_p(\tau_0)  \calR_p^*(\tau)   \big]  = \frac{-H^2 \tau_s^6}{12 M_P^2 \epsilon_i  h \tau_e^3 \tau^3 } \big( h \tau_e^3 + (6- h) \tau^3
 \big) \, ,
\end{equation}
and
\begin{equation}
 \mathrm{Im}  \big[  \calR_p(\tau_0)  \calR'^*_p(\tau)   \big]  = \frac{H^2 \tau_s^6}{4 M_P^2 \epsilon_i   \tau^4 }  \, .
\end{equation}
In particular, we observe that the above imaginary components are independent of $p$.

On the other hand, for the remaining imaginary components in $(a)$
 and $(e)$, the most dominant contribution scales like $p^{-6}$ in the limit $p \rightarrow 0$. This corresponds to the case where all factors of mode functions 
 $\calR_{p_i}$ are real and equal to its value at the end of inflation  $\calR_{p_i}(\tau_0)$:
 \ba
\calR_{p}(\tau) \simeq \calR_{p_i}(\tau_0) \simeq \frac{H}{2 \epsilon_i M_P \, p^{\frac{3}{2}}}  \, .
\ea 
This yields, 
\ba
\label{a-simple}
(a): 
-16 P_{\calR}(p_1)  P_{\calR}(p_2)  \mathrm{Im} \big[ \calR_{p}(\tau_0)  \calR^*_{p}(\tau_2)  \big] 
\mathrm{Im}\big[    \calR'_{q}(\tau_1)^2  {\calR_q'^*}(\tau_2)^2 \big]
+ 2\,  \mathrm{c. p.} 
\ea
and similarly for $(e)$ term:
\ba
\label{e-simple}
(e): 
-32 P_{\calR}(p_1)  P_{\calR}(p_2)  \mathrm{Im} \big[ \calR_{p}(\tau_0)  \calR'^*_{p}(\tau_2)  \big] 
\mathrm{Im}\big[    \calR'_{q}(\tau_1)^2  {\calR_q^*}(\tau_2) {\calR_q'^*}(\tau_2) \big]
+ 2\,  \mathrm{c. p.} 
\ea

Now we calculate ${\cal B}_{A_4 B_3}$ which is obtained from the contraction of the term with $A_4$ in ${\bf H_4}$ with the term containing  $B_3$ in ${\bf H_3}$, yielding 
\ba
\label{A4B3}
{\cal B}_{A_4 B_3}  =  \int_{-\infty}^{\tau_0} d \tau_2 \int_{-\infty}^{\tau_2} d \tau_1  A_4(\tau_1) B_3(\tau_2) 
\Big[ \prod_i^4 \int \frac{d^3 \bfq_i}{ (2 \pi)^\frac{3}{2}} (2 \pi)^3 \delta^3( \sum_i \bfq_i)\Big] 
\Big[\prod_j^3 \int \frac{d^3 \bfk_i}{(2 \pi)^\frac{3}{2}} (2 \pi)^3 \delta^3( \sum_i \bfk_i)
\Big]  \hspace{-1cm} \nonumber\\
 \times  \bigg \langle   \bigg[   \big( \hat\calR_{\bfq_1}   \hat\calR_{\bfq_2} \hat\calR'_{\bfq_3} \hat\calR'_{\bfq_4} \big) (\tau_1), \Big[
 \big( \hat\calR_{\bfp_1} \hat\calR_{\bfp_2} \hat\calR_{\bfp_3} \big) (\tau_0)  , \big( \hat\calR_{\bfk_1}  \hat\calR_{\bfk_2}   \hat\calR_{\bfk_3} \big) (\tau_2)  \Big]  \bigg]  \bigg \rangle i^2 k_3^2 \, .
\ea
As in the case of ${\cal B}_{A_4 A_3}$ there are 8 different contributions. The leading contribution is the one in which no contraction of $\hat\calR_{\bfp_i}$ with
 $ \partial^2 \hat\calR_{\bfk_i}$ is made. 
 Considering the second line involving the contractions and following the same logic as the terms $(a)$ above, the leading contribution containing order $p^{-6}$ is given by
 \ba
 \label{a1-simple} 
-32 i^2 q^2 P_{\calR}(p_1)  P_{\calR}(p_2)  \mathrm{Im} \big[ \calR_{p}(\tau_0)  \calR^*_{p}(\tau_2)  \big] 
\mathrm{Im}\big[    \calR'_{q}(\tau_1)^2  {\calR_q^*}(\tau_2)^2 \big]
+ 2\,  \mathrm{c. p.}  
\ea 

On the other hand, the term  ${\cal B}_{B_4 A_3}$ has the following form 
\ba
\label{B4A3}
{\cal B}_{B_4 A_3}  =  \int_{-\infty}^{\tau_0} d \tau_2 \int_{-\infty}^{\tau_2} d \tau_1  B_4(\tau_1) A_3(\tau_2) 
\Big[ \prod_i^4 \int \frac{d^3 \bfq_i}{ (2 \pi)^\frac{3}{2}} (2 \pi)^3 \delta^3( \sum_i \bfq_i)\Big] 
\Big[\prod_j^3 \int \frac{d^3 \bfk_i}{(2 \pi)^\frac{3}{2}} (2 \pi)^3 \delta^3( \sum_i \bfk_i)
\Big]  \hspace{-1cm} \nonumber\\
 \times  \bigg \langle   \bigg[   \big( \hat\calR_{\bfq_1}   \hat\calR_{\bfq_2} \hat\calR_{\bfq_3} \hat\calR_{\bfq_4} \big) (\tau_1), \Big[
 \big( \hat\calR_{\bfp_1} \hat\calR_{\bfp_2} \hat\calR_{\bfp_3} \big) (\tau_0)  , \big( \hat\calR_{\bfk_1}  \hat\calR'_{\bfk_2}   \hat\calR'_{\bfk_3} \big) (\tau_2)  \Big]  \bigg]  \bigg \rangle i^2 {\bfq_3}\cdot {\bfq_4} \, .
\ea
Following the same steps as in the case of ${\cal B}_{A_4 A_3}$, there are two leading contributions of order $p^{-6}$ as follows:
\ba
\label{a2-simple}
16 i^2 q^2 P_{\calR}(p_1)  P_{\calR}(p_2)  \mathrm{Im} \big[ \calR_{p}(\tau_0)  \calR^*_{p}(\tau_2)  \big] 
\mathrm{Im}\big[    \calR_{q}(\tau_1)^2  {\calR_q'^*}(\tau_2)^2 \big]
+ 2\,  \mathrm{c. p.} 
\ea
and
\ba
\label{e2-simple} 
32 i^2 q^2 P_{\calR}(p_1)  P_{\calR}(p_2)  \mathrm{Im} \big[ \calR_{p}(\tau_0)  \calR'^*_{p}(\tau_2)  \big] 
\mathrm{Im}\big[    \calR_{q}(\tau_1)^2  {\calR_q'^*}(\tau_2) {\calR_q^*}(\tau_2) \big]
+ 2\,  \mathrm{c. p.} 
\ea

Finally, ${\cal B}_{B_4 A_3}$ has the following form 
\ba
\label{B4B3}
{\cal B}_{B_4 B_3}  =  \int_{-\infty}^{\tau_0} d \tau_2 \int_{-\infty}^{\tau_2} d \tau_1  B_4(\tau_1) B_3(\tau_2) 
\Big[ \prod_i^4 \int \frac{d^3 \bfq_i}{ (2 \pi)^\frac{3}{2}} (2 \pi)^3 \delta^3( \sum_i \bfq_i)\Big] 
\Big[\prod_j^3 \int \frac{d^3 \bfk_i}{(2 \pi)^\frac{3}{2}} (2 \pi)^3 \delta^3( \sum_i \bfk_i)
\Big]  \hspace{-1cm} \nonumber\\
 \times  \bigg \langle   \bigg[   \big( \hat\calR_{\bfq_1}   \hat\calR_{\bfq_2} \hat\calR_{\bfq_3} \hat\calR_{\bfq_4} \big) (\tau_1), \Big[
 \big( \hat\calR_{\bfp_1} \hat\calR_{\bfp_2} \hat\calR_{\bfp_3} \big) (\tau_0)  , \big( \hat\calR_{\bfk_1}  \hat\calR_{\bfk_2}   \hat\calR_{\bfk_3} \big) (\tau_2)  \Big]  \bigg]  \bigg \rangle i^4\,  k_3^2 \,  {\bfq_3}\cdot {\bfq_4} \, .
\ea
It has the following leading contribution in its contractions: 
\ba
\label{a3-simple}
32 i^4 q^4 P_{\calR}(p_1)  P_{\calR}(p_2)  \mathrm{Im} \big[ \calR_{p}(\tau_0)  \calR^*_{p}(\tau_2)  \big] 
\mathrm{Im}\big[    \calR_{q}(\tau_1)^2  {\calR_q^*}(\tau_2)^2 \big]
+ 2\,  \mathrm{c. p.} 
\ea

Combining the leading terms (\ref{a3-simple}),  (\ref{a2-simple}), (\ref{a1-simple}) and (\ref{a-simple}) which all have common imaginary factor 
$\mathrm{Im} \big[ \calR_{p}(\tau_0)  \calR^*_{p}(\tau_2)  \big] $, 
yields the following contribution in ${\cal B}'$:
\ba
\label{B-1}
-16  P_{\calR}(p_1)  P_{\calR}(p_2) \int_{-\infty}^{\tau_0} d \tau_2 \int_{-\infty}^{\tau_2} d \tau_1 
\mathrm{Im} \big[ \calR_{p}(\tau_0)  \calR^*_{p}(\tau_2)  \big]  
\int \frac{d^3 \bfq}{ (2 \pi)^{3}} 
\mathrm{Im} \big[  I_4(q, \tau_1) I_3(q, \tau_2)\big] \, ,
\ea
in which the quantities $I_4(q, \tau)$ and $I_3(q, \tau)$ are given by:
\ba
I_4(q, \tau) \equiv  
 \Big[ A_4 (\tau) \calR_q'(\tau)^2  + q^2 B_4(\tau)  \calR_q(\tau)^2 \Big]  \, ,
\ea
and
\ba
I_3(q, \tau) \equiv  \Big[ A_3 (\tau) \calR_q^{'*}(\tau)^2  -2 q^2 B_3(\tau)  \calR_q^*(\tau)^2 \Big] \, .
\ea 
On the other hand, combining the remaining two terms (\ref{e-simple}) and (\ref{e2-simple}) which both have the common imaginary factor 
$\mathrm{Im} \big[ \calR_{p}(\tau_0)  \calR'^*_{p}(\tau_2)  \big] $, 
yields the following contribution in ${\cal B}'$:
\ba
\label{B-2}
-32  P_{\calR}(p_1)  P_{\calR}(p_2) \int_{-\infty}^{\tau_0} d \tau_2 \int_{-\infty}^{\tau_2} d \tau_1 
\mathrm{Im} \big[ \calR_{p}(\tau_0)  \calR'^*_{p}(\tau_2)  \big]   A_3(\tau_2) 
\int \frac{d^3 \bfq}{ (2 \pi)^{3}}  
\mathrm{Im} \big[  I_4(q, \tau_1)  \calR_q^{'*}(\tau_2) \calR_q^{*}(\tau_2) \big] . 
\ea
Finally, combining the above results (\ref{B-1}) and (\ref{B-2}) yields the total contributions for ${\cal B}$ as given in Eq. (\ref{B-result}). 

The analysis for ${\cal C}$ goes similar as above, with the difference being that the positions of ${\bf H}_3$ and ${\bf H}_4$ are switched. Here we provide some steps yielding to ${\cal C}_{A_3 A_4}$ which is given by 
\ba
\label{C-AA}
{\cal C}_{A_3 A_4}  =  \int_{-\infty}^{\tau_0} d \tau_2 \int_{-\infty}^{\tau_2} d \tau_1  A_3(\tau_1) A_4(\tau_2) 
\Big[ \prod_i^4 \int \frac{d^3 \bfq_i}{ (2 \pi)^\frac{3}{2}} (2 \pi)^3 \delta^3( \sum_i \bfq_i)\Big] 
\Big[\prod_j^3 \int \frac{d^3 \bfk_i}{(2 \pi)^\frac{3}{2}} (2 \pi)^3 \delta^3( \sum_i \bfk_i)
\Big]  \hspace{-1cm} \nonumber\\
 \times  \bigg \langle   \bigg[  \big( \hat\calR_{\bfk_1}  \hat\calR'_{\bfk_2}   \hat\calR'_{\bfk_3} \big) (\tau_1) , \Big[
 \big( \hat\calR_{\bfp_1} \hat\calR_{\bfp_2} \hat\calR_{\bfp_3} \big) (\tau_0)  ,   \big( \hat\calR_{\bfq_1}   \hat\calR_{\bfq_2} \hat\calR'_{\bfq_3} \hat\calR'_{\bfq_4} \big) (\tau_2) \Big]  \bigg]  \bigg \rangle \, .
\ea
As in the case of ${\cal B}_{A_4 A_3}$, there are 8 different contributions into the above contraction. As in ${\cal B}_{A_4 A_3}$, only 2 contributions similar to $(a)$ term and $(e)$ term are leading. The term similar to the $(a)$ term is given by 
\ba
-16 \mathrm{Im} \big[ \calR_{p_1}(\tau_0)  \calR_{p_2}(\tau_0)   
\calR^*_{p_1}(\tau_2)  \calR^*_{p_2}(\tau_2)  \big] 
\mathrm{Im}\big[    \calR'_{q}(\tau_1)^2  {\calR_q'^*}(\tau_2)^2 
\calR_{p_3}(\tau_1)  \calR^*_{p_3}(\tau_0) \big]
+ 2\,  \mathrm{c. p.} 
\ea
From the above expression, one can show that the leading order contribution containing  $p^{-6}$ is given by 
\ba
\label{a-Cterm}
-16  \big( P_{\calR}(p_1)  + P_{\calR}(p_2) \big)   P_{\calR}(p_3) 
\mathrm{Im} \big[ \calR_{p}(\tau_0)  \calR^*_{p}(\tau_2)  \big] 
\mathrm{Im}\big[    \calR'_{q}(\tau_1)^2  {\calR_q'^*}(\tau_2)^2 \big]
+ 2\,  \mathrm{c. p.} 
\ea 
On the other hand, the term similar to the $(e)$ term in ${\cal B}_{A_4 A_3}$ has the following form:
\ba
-32 \mathrm{Im} \big[ \calR_{p_1}(\tau_0)  \calR_{p_2}(\tau_0)   
\calR'^*_{p_1}(\tau_2)  \calR^*_{p_2}(\tau_2) + p_1 \leftrightarrow p_2 \big] 
\mathrm{Im}\big[    \calR'_{q}(\tau_1)^2  {\calR_q'^*}(\tau_2)  {\calR_q^*}(\tau_2) 
\calR_{p_3}(\tau_1)  \calR^*_{p_3}(\tau_0) \big]
+ 2\,  \mathrm{c. p.} \nonumber 
\ea
This is further simplified into 
\ba
\label{e-Cterm}
-32  \big( P_{\calR}(p_1)  + P_{\calR}(p_2) \big)   P_{\calR}(p_3) 
\mathrm{Im} \big[ \calR_{p}(\tau_0)  \calR'^*_{p}(\tau_2)  \big] 
\mathrm{Im}\big[    \calR'_{q}(\tau_1)^2  {\calR_q'^*}(\tau_2) 
{\calR_q^*}(\tau_2)  \big]
+ 2\,  \mathrm{c. p.} 
\ea 
Combining Eqs. (\ref{e-Cterm}) and (\ref{a-Cterm}) yields the following expression for ${\cal C}_{A_3 A_4}$:
\ba
\label{CA3A4}
&&{\cal C}_{A_3 A_4} = 
-32  P_{\calR}(p_1)  P_{\calR}(p_2)  \int \frac{d^3 \bfq}{ (2 \pi)^{3}} 
\int_{-\infty}^{\tau_0} d \tau_2 \int_{-\infty}^{\tau_2} d \tau_1 
A_3( \tau_1) A_4(\tau_2)  \times  \\
&&\bigg\{\mathrm{Im} \big[ \calR_{p}(\tau_0)  \calR^*_{p}(\tau_2)  \big]  
\mathrm{Im}\big[    {\calR'_{q}}^2(\tau_1)  {\calR_q'^*}^2(\tau_2) \big]
+ \mathrm{Im} \big[ \calR_{p}(\tau_0)  \calR'^*_{p}(\tau_2)  \big] 
\mathrm{Im}\big[    {\calR'_{q}}^2(\tau_1)  {\calR_q'^*}(\tau_2) 
{\calR_q^*}(\tau_2)  \big] \bigg\} 
+ 2  \mathrm{c. p.}\nonumber
\ea
Calculating similarly ${\cal C}_{A_3 B_4}, {\cal C}_{B_3 A_4}$ and ${\cal C}_{B_3 B_4}$ yield our final expression for ${\cal C}$ as given in Eq. (\ref{C-result}). 
  
 \vspace{0.5cm}
  


{}

\end{document}